\documentclass[twocolumn,twocolappendix,tighten]{aastex631}
\usepackage{bm}

\received{2022 Sep 7}
\revised{2023 Mar 31}
\accepted{2023 May 30}
\published{DATE}

\graphicspath{{./}{figures/}}

\newcommand{\zsun}{Z_\odot}

\newcommand{\msun}{M_\odot}
\newcommand{\msunyr}{M_\odot\,{\rm yr}^{-1}}
\newcommand{\nh}{n_{\rm H}}
\newcommand{\cc}{{\rm cm^{-3}}}

\shorttitle{Metal-enriched Atomic-cooling Halos}
\shortauthors{Hirano, Machida, and Basu}

\begin{document}

\title{Magnetic Effects Promote Supermassive Star Formation in Metal-enriched Atomic-cooling Halos}

\correspondingauthor{Shingo Hirano}
\email{hirano@astron.s.u-tokyo.ac.jp}

\author[0000-0002-4317-767X]{Shingo Hirano}
\affiliation{Department of Astronomy, School of Science, University of Tokyo, Tokyo 113-0033, Japan}
\affiliation{Department of Earth and Planetary Sciences, Faculty of Science, Kyushu University, Fukuoka 819-0395, Japan}

\author[0000-0002-0963-0872]{Masahiro N. Machida}
\affiliation{Department of Earth and Planetary Sciences, Faculty of Science, Kyushu University, Fukuoka 819-0395, Japan}
\affiliation{Department of Physics and Astronomy, University of Western Ontario, London, ON N6A 3K7, Canada}

\author[0000-0003-0855-350X]{Shantanu Basu}
\affiliation{Department of Physics and Astronomy, University of Western Ontario, London, ON N6A 3K7, Canada}

\begin{abstract}
Intermediate-mass black holes (with $\geq\!10^5\,\msun$) are promising candidates for the origin of supermassive black holes (with $\sim\!10^9\,\msun$) in the early universe (redshift $z\sim6$).
\cite{chon20} firstly pointed out the direct collapse black hole (DCBH) formation in metal-enriched atomic-cooling halos (ACHs), which relaxes the DCBH formation criterion.
On the other hand, \cite{hirano21} showed that the magnetic effects promote the DCBH formation in metal-free ACHs.
We perform a set of magnetohydrodynamical simulations to investigate star formation in the magnetized ACHs with metallicities $Z/\zsun = 0$, $10^{-5}$, and $10^{-4}$.
Our simulations show that the mass accretion rate onto the protostars becomes lower in metal-enriched ACHs than that of metal-free ACHs.
However, many protostars form from gravitationally and thermally unstable metal-enriched gas clouds. 
Under such circumstances, the magnetic field rapidly increases as the magnetic field lines wind up due to the spin of protostars.
The region with the amplified magnetic field expands outwards due to the orbital motion of protostars and the rotation of the accreting gas.
The amplified magnetic field extracts the angular momentum from the accreting gas, promotes the coalescence of the low-mass protostars, and increases the mass growth rate of the primary protostar.
We conclude that the magnetic field amplification is always realized in the metal-enriched ACHs regardless of the initial magnetic field strength, which affects the DCBH formation criterion.
In addition, we find a qualitatively different trend from the previous unmagnetized simulations in that the mass growth rate is maximal for the extremely metal-poor ACHs with $Z/\zsun = 10^{-5}$.
\end{abstract}


\section{Introduction} \label{sec:intro}

The formation of the supermassive black holes (SMBHs) with mass $\sim\!10^9\,\msun$ in the early universe at redshift $z\geq6$ is one of the outstanding issues in astrophysics \citep[see][as review]{woods19}.
As researchers find SMBHs in earlier epochs of the universe, the demands on theoretical models become tighter.
The number of observational samples will increase with upcoming observations, such as with the James Webb Space Telescope (JWST).
The theoretical formation scenario for SMBHs should be updated accordingly.

\begin{figure}[t]
\begin{center}
\includegraphics[width=1.0\columnwidth]{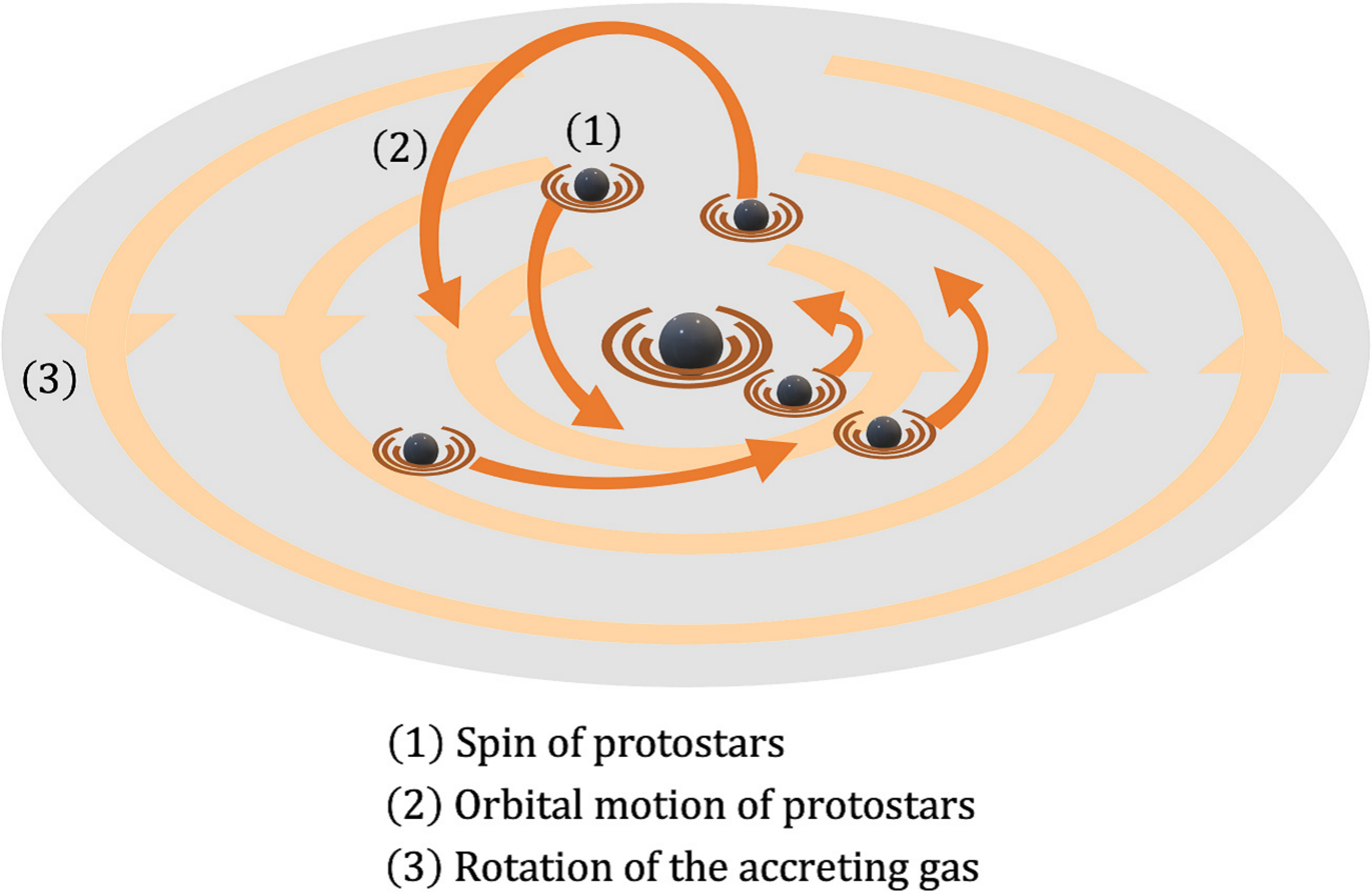}
\end{center}
\caption{
Schematic overview of the rotation-driven magnetic field amplification after the protostar formation.
(1) The spin motion of each protostar amplifies the magnetic fields around them.
(2) The amplified ``seed'' magnetic field enhances the surrounding magnetic fields by the orbital motion of protostars. 
(3) The rotation of the accreting gas gradually expands the amplified field region outwards.
We note that the magnetic field amplification would not operate in the present-day star formation because the magnetic field significantly dissipates within the disk.
}
\label{f1}
\end{figure}

One of the popular formation channels of SMBHs is the direct collapse black hole (DCBH) scenario \citep[see][as review]{inayoshi20}.
In the metal-free atomic-cooling halos (ACHs), the gas component is cooled only by the atomic hydrogen and can gravitationally contract while remaining at a high temperature $\sim\!8000\,$K.
The high temperature leads to a large Jeans mass, suppression of cloud fragmentation, and also a high mass accretion rate that allows the protostar to increase its mass efficiently.
The high mass accretion rate leads to an expanded envelope of the protostar with a relatively low surface temperature of a few times $1000\,$K \citep{hosokawa12}, hence limiting radiative feedback that can otherwise halt accretion.
The result is the formation of a supermassive star (SMS).
Once the SMS exceeds the mass threshold of the general relativistic instability, the star can collapse to form a massive black hole with a similar mass \citep[$\sim\!10^5\,\msun$;][]{umeda16}\footnote{Recently, \cite{nagele22arXiv} report a new mass range for general relativistic instability supernovae as $2.6$--$3.6\times10^4\,\msun$. If the metallicity is low enough, mass loss during stellar evolution is negligible (a few percent of the initial stellar mass), so the final fate of metal-enriched SMS is expected to be almost the same as that of metal-free stars.}.
This marks the formation of an intermediate-mass black hole (IMBH) with $\sim\!10^5\,\msun$, a candidate seed of SMBHs.
To explain the observed number density of the high-redshift quasars \citep[a few Gpc$^{-3}$; e.g.,][]{banados16}, the previous studies provided a number of conditions under which the metal-free ACHs can be realized: H$_2$-dissociating ultraviolet radiation \citep[e.g.,][]{omukai01,agarwal12,latif13}, high-velocity collisions \citep{inayoshi15}, baryon-dark matter streaming velocities \citep{tanaka14,hirano17sv}, dynamical heating due to the violent halo merger \citep{wise19}, and turbulent cold flows \citep{latif22}.

Recently, \cite{chon20} pointed out the possibility of IMBH formation in metal-enriched ACHs.
Traditionally, researchers have considered that the low-metallicity gas clouds were unsuitable for the IMBH formation because the low gas temperature due to the metal and dust cooling produce numerous low-mass protostars, reducing the accretion rate onto each protostar by fragmentation.
\cite{chon20} performed a set of hydrodynamical simulations to follow the star formation in ACHs with different metallicities $Z/\zsun = 10^{-6}$--$10^{-3}$.
They found that for slightly metal-enriched ACHs with $Z/\zsun \leq 10^{-4}$, thousands of stars are formed by frequent fragmentation of the gas cloud.
However, most stars competitively merge into the most massive (primary) star (``supercompetitive accretion'').
The mass growth rate of the primary star in metal-enriched ACHs is almost the same as in metal-free ACHs.
This relaxes the criterion of the DCBH formation, from $Z/\zsun = 0$ to $\leq 10^{-4}$.
Allowing DCBH formation at metal-enriched ACHs increases the DCBH formation rate compared to the case with only metal-free ACHs.

This study investigates whether the magnetic effects can support the DCBH formation in the metal-enriched ACHs.
In the metal-free cases, we reported that the cosmological seed magnetic field is efficiently amplified during the early accretion phase, as shown in Figure~\ref{f1} \citep{hirano22}.
The resultant strong magnetic field promotes the coalescence of protostars and enhances the mass accretion onto the primary protostar at the center of the metal-free ACHs \citep[][hereafter Paper I]{hirano21}.
Due to the cosmic time evolution, the metal-enriched ACHs may initially have a stronger magnetic field than the metal-free ACHs.
Therefore, we speculate that the magnetic effects might be more effective for the metal-enriched ACHs than for the metal-free ACHs.
This paper investigates whether or not the magnetic effects can promote the DCBH formation in metal-enriched ACHs.

\section{Numerical Methodology} \label{sec:methods}

\begin{figure}[t]
\begin{center}
\includegraphics[width=1.0\columnwidth]{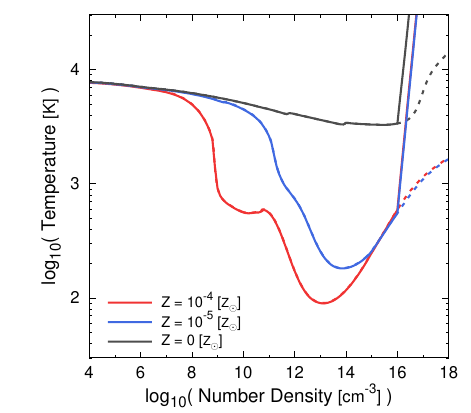}
\end{center}
\caption{
Thermal evolution models of the atomic-cooling halos with different metallicity, $Z/\zsun = 0$, $10^{-5}$, and $10^{-4}$, respectively, as a function of the gas number density.
The dashed lines are the base model theoretically obtained from the chemical reaction calculation under the strong Lyman-Werner radiation with $J_{21} = 10^3$ \citep{omukai08,machida15}.
The solid lines, the adopted models in this study, are variants of the dashed lines using the stiff-EOS technique \citep[see][]{hirano22} with threshold density $n_{\rm th} = 10^{16}\,\cc$.
}
\label{f2}
\end{figure}

We perform a set of three-dimensional (3D) ideal magnetohydrodynamical (MHD) simulations of the star formation in the metal-enriched ACHs.
The numerical simulation setup follows Paper I, but we switch the thermal evolution models of the ACHs according to the metallicities, as shown in Figure~\ref{f2}.

\begin{deluxetable}{ccccc}[t]
\tablecaption{Parameters of Initial Clouds and Calculation Results}
\tablecolumns{12}
\tablenum{1}
\tablewidth{0pt}
\tablehead{
  \colhead{1} &
  \colhead{2} &
  \colhead{3} &
  \colhead{4} &
  \colhead{5} \\
  \colhead{Model} &
  \colhead{$Z$} &
  \colhead{$B_0$} &
  \colhead{$t_{\rm col}$} &
  \colhead{$t_{\rm ps}$} \\
  \colhead{} &
  \colhead{($Z_\odot$)} &
  \colhead{(G)} &
  \colhead{(yr)} &
  \colhead{(yr)}
}
\startdata
Z4B00 & $10^{-4}$ & 0          & 1,177,683 & 500 \\
Z4B12 &           & $10^{-12}$ &       -34 & 500 \\
Z4B10 &           & $10^{-10}$ &       -39 & 250 \\
Z4B08 &           & $10^{-8}$  &       -15 & 500 \\
Z4B06 &           & $10^{-6}$  &       -38 & 290 \\
\hline
Z5B00 & $10^{-5}$ & 0          & 1,206,806 & 500 \\
Z5B12 &           & $10^{-12}$ &        +8 & 300 \\
Z5B10 &           & $10^{-10}$ &       -49 & 230 \\
Z5B08 &           & $10^{-8}$  &       -91 & 280 \\
Z5B06 &           & $10^{-6}$  &       +34 & 210 \\
\hline
Z0B00 & $0 \simeq 10^{-6}$ & 0 & 1,209,600 & 500 \\
Z0B12 &           & $10^{-12}$ &       -98 & 500 \\
Z0B10 &           & $10^{-10}$ &       -72 & 500 \\
Z0B08 &           & $10^{-8}$  &      -108 & 360 \\
Z0B06 &           & $10^{-6}$  &       -98 & 190 \\
\enddata
\tablecomments{
Column (1): model name.
Columns (2) and (3): parameters $Z$ (the metallicity) and $B_0$ (the initial magnetic field strength) of the initial cloud.
Column (4): elapsed time when the first protostar forms (corresponding to $t_{\rm ps} = 0\,$yr).
Column (5): elapsed time at the end of the simulation after the first protostar formation.
}
\label{t1}
\end{deluxetable}

\subsection{Numerical methodology} \label{sec:methodology}

We solve the ideal MHD\footnote{We assume that the non-ideal MHD effects are ineffective in the metal-enriched ACHs with high temperature $>\!100\,$K in this study, but discuss it in Section~\ref{sec:dis-caution}.} equations \cite[equations~(1)--(4) of][]{machida13} with the barotropic equation of state (EOS) tables.
We adopt EOS tables of ACHs under the strong Lyman-Werner radiation\footnote{$J_{21}$ is the intensity at the Lyman-Werner bands normalized in units of $10^{-21}$\,erg\,s$^{-1}$\,cm$^{-2}$\,Hz$^{-1}$\,sr$^{-1}$., $J_{21} = 10^3$} for a metallicity with $Z/\zsun = 0$, $10^{-5}$, and $10^{-4}$ based on a chemical reaction simulation \citep{omukai08} as used in \cite{chon20}.
We assume a blackbody spectrum of Population II stellar sources with the effective temperature of $10^4$\,K to realize the strong Lyman-Werner radiation.
We define the EOS model with $Z/\zsun = 10^{-6}$ as the metal-free case ($Z/\zsun = 0$) because the thermal evolution with $Z/\zsun = 0$ follows that with $Z/\zsun = 10^{-6}$ \citep{omukai08}.
Instead of a sink particle technique, we adopt a stiff-EOS technique to accelerate the time evolution while connecting magnetic field lines to high-density regions, which is necessary to reproduce the magnetic field amplification \citep[Paper~I,][]{hirano22}.
We set a threshold density $n_{\rm th} = 10^{16}\,\cc$, which reproduces dense cores whose radius is roughly consistent with the approximate mass--radius relation of a rapidly accreting protostar with $R_* \simeq 12 (M_*/100\,\msun)^{1/2}\,$au \citep[Equation~(11) of][]{hosokawa12}. 
Note that the difference in the radius between our model and \citet{hosokawa12} is about 10 at the maximum. The radii of the formed protostar in our simulations tend to be smaller than those in \citet{hosokawa12}.
Figure~\ref{f2} shows the resultant EOS tables.

We use our nested grid code \citep{machida15}, in which the rectangular grids of ($n_x$, $n_y$, $n_z$) = ($256$, $256$, $32$) are superimposed.
The grid size $L(l)$ and cell width $h(l)$ of the $l$th grid are twice larger than those of ($l+1$)th grid (e.g., $L(l) = 2L(l+1)$ and $h(l) = 2h(l+1)$).
We use the index ``$l$'' to describe a grid level. 
The base grid ($l = 1$) has the grid size $L(1) = 6.68\times10^7\,$au and the cell size $h(1) = 2.61\times10^5\,$au, respectively.
A new finer grid is generated to resolve the Jeans wavelength of at least 32 cells.
The finest grid ($l = 18$) has the grid size $L(18) = 510\,$au and the cell size $h(18) = 1.99\,$au, respectively.

\subsection{Initial Condition}

We adopt the initial condition as a cloud with an enhanced Bonnor-Ebert (BE) density profile $f \cdot n_{\rm BE}(r)$ where $f = 1.2$ is an enhanced factor to promote the cloud contraction.
The initial cloud has a central density $n_{0}(0) = f \cdot n_{\rm BE}(0) = 1.2 \times 10^4\,\cc$, mass $M_0 = 1.82\times10^{6}\,\msun$, and radius $R_0 = 2.09\times10^6$\,au = $10.1$\,pc, respectively.
We impose a uniform temperature $T_0 = 7700\,$K and a rigid rotation of $\Omega_0 = 7.67 \times 10^{-15}\,{\rm s^{-1}}$ for the initial cloud.
The resultant ratio of thermal and rotational energies to the gravitational energy of the initial cloud are $\alpha_0 = 0.7$ and $\beta_0 = 0.01$, respectively.
We add an $m = 2$ mode nonaxisymmetric density perturbation to the initial cloud with the amplitude of the perturbation as 10\%.
We do not include turbulence and do not consider a small-scale dynamo \citep[e.g.,][]{sur10,mckee20} because we only consider very weak fields which are significantly amplified by the rotation motion of protostars and accreting gas (Figure~\ref{f1}).

\subsection{Model Parameters}

We adopt two parameters for the gas clouds: (1) metallicities and (2) initial magnetic field strength.
We adopt three different gas metallicities $Z/\zsun = 0$, $10^{-5}$, and $10^{-4}$, which cover the metallicity range for the ``supercompetitive accretion'' ($Z/ \zsun \leq10^{-4}$; \citealt{chon20}). 
We adopt five different initial magnetic field strengths $B_0/{\rm G} = 0$, $10^{-12}$, $10^{-10}$, $10^{-8}$, and $10^{-6}$.
We adopt the magnetic field strength $B_0/{\rm G} = 10^{-12}$ at $\nh = 10^4\,\cc$ extrapolated from the cosmological seed value \citep[$10^{-15}$\,G at $\nh = 1\,\cc$;][]{xu08} through flux freezing during the cloud compression (described by a power law $B \propto n_{\rm H}^{2/3}$) as the minimum value.
We impose a uniform magnetic field with the same direction as the initial cloud's rotation axis in the whole computational domain.
In Cartesian coordinates, the initial directions of the global magnetic field and the rotation axis are parallel to the $z$-axis.

Table~\ref{t1} summarizes the simulation models adopted in this study.
We define the model names by connecting the common logarithms of two model parameters.
We adopt model Z4B12 with ($Z/\zsun$, $B_0$/G) = ($10^{-4}$, $10^{-12}$) as the fiducial model.
We expect that it is the most difficult for model Z4B12 to form SMS due to the lowest accretion rate (corresponding to the lowest gas temperature) and the weakest magnetic field among the models in this study.
We can conclude that the magnetic effects support DCBH formation in any other model by confirming the formation of an SMS in model Z4B12.

\subsection{Simulations}

We assume that the region where the gas density exceeds $n_{\rm th}$ is a dense core that hosts a protostar.
We analyze the number and masses of such dense cores to examine the fragmentation process.
We define the epoch of the first protostar formation ($t_{\rm ps} = 0\,$yr) when the gas number density firstly reached the threshold density ($n_{\rm max} = n_{\rm th}$).
We terminate the calculation either when $t_{\rm ps} = 500\,$yr or the calculation timestep becomes very short ($dt\ll0.01\,$yr) because we cannot follow the further time evolution of the system.
We could calculate the time evolution for about $200$--$500\,$yr after the first protostar formation (Table~\ref{t1}).

\begin{figure*}[t]
\begin{center}
\includegraphics[width=1.0\linewidth]{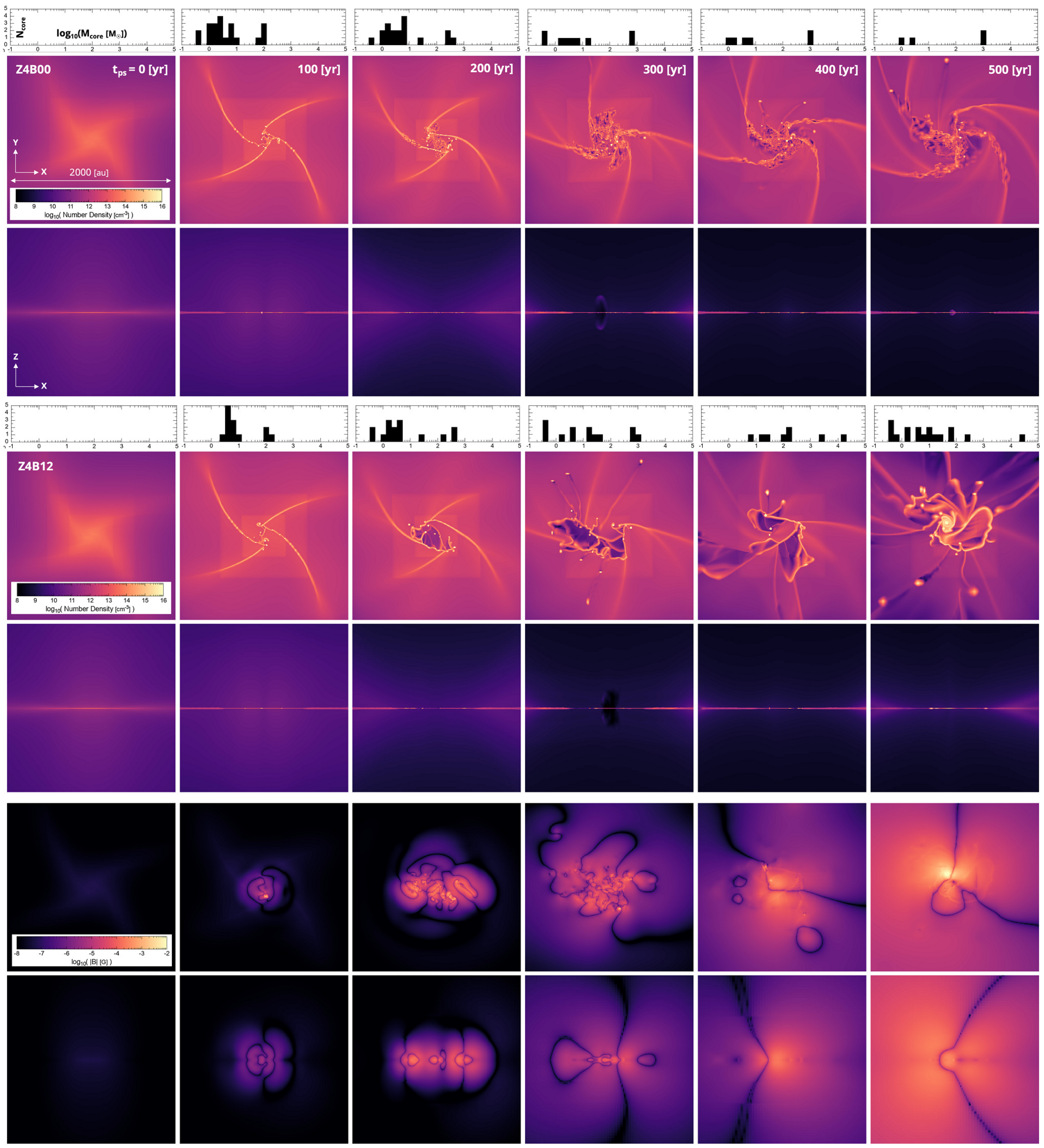}
\end{center}
\caption{
Mass spectrum of dense cores (protostars) and cross-sectional views on the $z = 0$ and $y = 0$ planes around the most massive protostar at $t_{\rm ps} = 0$, $100$, $200$, $300$, $400$, and $500\,$yr after the first protostar formation, respectively (from left to right panels).
The cross-sectional views are gas number density for models Z4B00 and Z4B12, and absolute magnetic field strength for model Z4B12.
The box sizes of the cross-sectional views are $2000\,$au.
The dark ``nodes'' seen in the magnetic field strength distribution are boundaries where the magnetic field direction reverses and the absolute magnetic field strength is very small.
}
\label{f3}
\end{figure*}

\begin{figure*}[t]
\begin{center}
\includegraphics[width=0.95\linewidth]{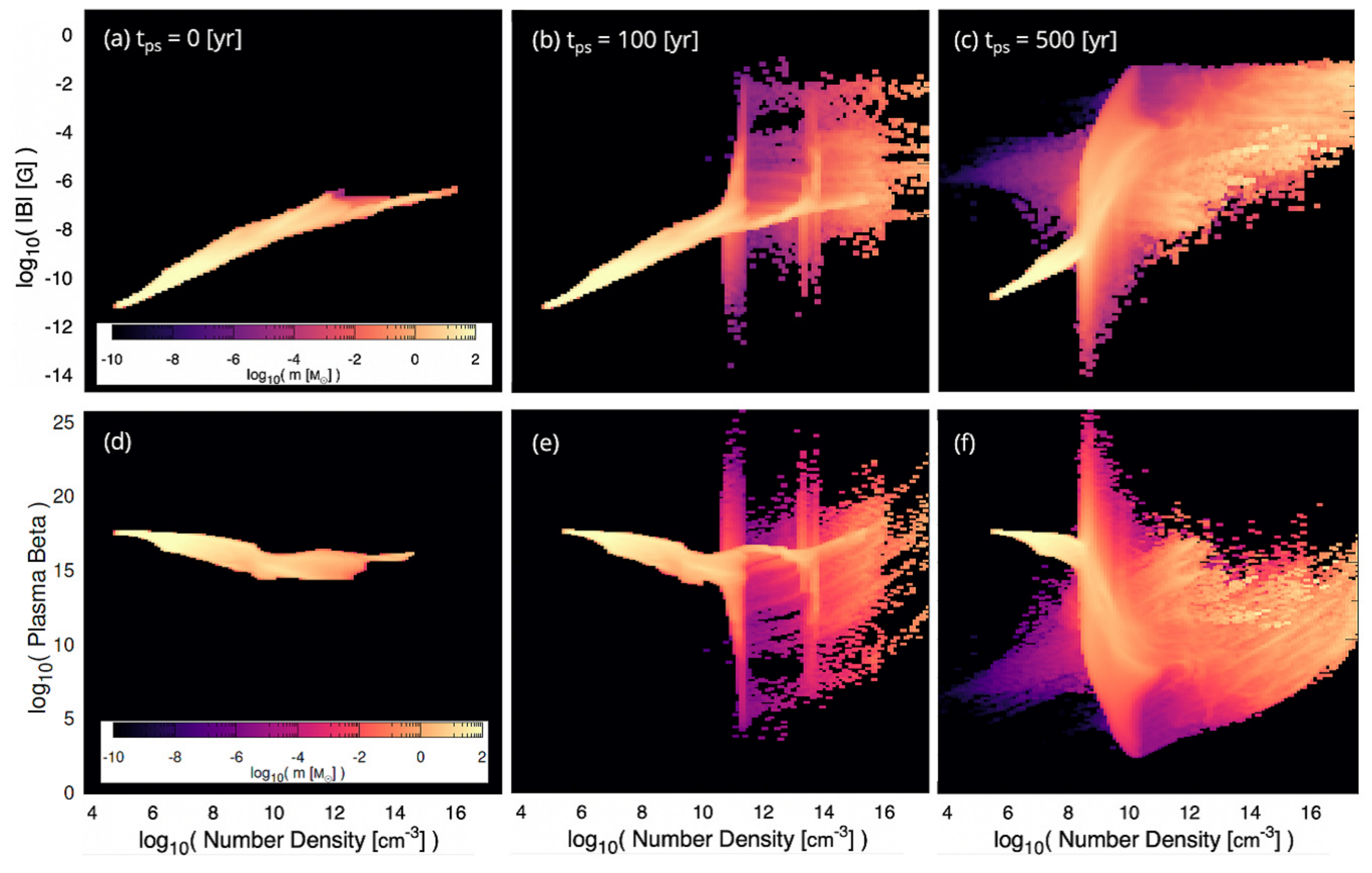}
\end{center}
\caption{
Phase diagrams of the absolute magnetic field strength (top panels) and plasma beta (bottom panels) for model Z4B12 (with $Z/\zsun = 10^{-4}$ and $B_0/{\rm G} = 10^{-12}$) at $t_{\rm ps} = 0$, $100$, and $500\,$yr after the first protostar formation (from left to right panels).
}
\label{f4}
\end{figure*}

\section{Results} \label{sec:res}

This section shows the dependence of the simulation results on two parameters.
Firstly, Section~\ref{sec:res-fiducial} presents the simulation results of the fiducial model (Z4B12) and compares it with the corresponding non-magnetized model (Z4B00).
Next, Section~\ref{sec:res-Z} confirms the dependence of the magnetic effects on the metallicities ($Z$).
Finally, Section~\ref{sec:res-B0} shows the dependence on the initial magnetic field strength ($B_0$).
In all metal-enriched models, we confirm the magnetic field amplification as in the metal-free ACHs and the increase of the mass accretion rate onto the primary protostar due to the angular momentum transfer by magnetic effects.

\subsection{Fiducial model} \label{sec:res-fiducial}

Figure~\ref{f3} compares the simulation results of the fiducial model (Z4B12) and the unmagnetized model (Z4B00) during the first $500\,$yr of the protostar accretion phase.
We place three different panels at each epoch: mass spectrum of the dense cores ($N_{\rm core}$)\footnote{As the calculation progresses, several small (low-mass) cores are ejected from the center and move to the region with coarse numerical resolution. Such cores are dissolved numerically. We note that the number of the low-mass cores is underestimated in the mass spectrum.}, maps of the gas number density distributions on the $z = 0$ and $y = 0$ planes ($n$), and maps of the absolute magnetic field strength on the $z = 0$ plane ($|B|$).
When the first protostar forms at the center of the collapsing gas cloud ($t_{\rm ps} = 0\,$yr), the gas density maps are identical between models Z4B12 and Z4B00 because the magnetic field strength in the vicinity of the protostar is too weak ($|B|\sim10^{-7}\,$G at most as Figure~\ref{f4}(a)) to affect the dynamics of the collapsing gas cloud.

\begin{figure}[t]
\begin{center}
\includegraphics[width=1.0\linewidth]{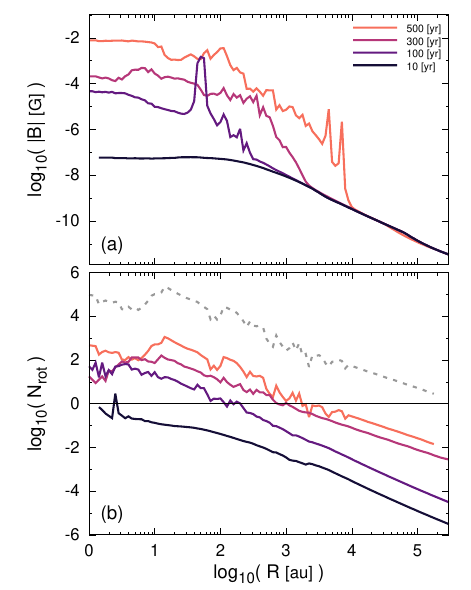}
\end{center}
\caption{
Radial profiles around the primary core for model Z4B12 at $t_{\rm ps} = 10$, $100$, $300$, and $500\,$yr after the first protostar formation.
Panels: (a) magnetic field strength and (b) number of orbital rotations $N_{\rm rot} = (v_{\rm rot} t_{\rm ps})/(2 \pi R)$ at $t_{\rm ps}$.
The dotted line in panel (b) represents the expected $N_{\rm rot}$ at $t_{\rm ps} = 10^5\,$yr when the first star ends its accretion phase \citep[Figure~1 in][]{hirano17}, by using the radial profile of rotational velocity at $t_{\rm ps} = 500\,$yr.
The horizontal line indicates $N_{\rm rot} = 1$.
Note that physical quantities are azimuthally averaged.
}
\label{f5}
\end{figure}

During the first $100\,$yr after the first protostar formation, the region around the cloud's center becomes gravitationally unstable due to the high mass accretion rate.
Fragmentation occurs and forms many protostars (see the mass spectrum in Figure~\ref{f3}).
The rapid spin of protostars winds up magnetic fields and amplifies the local magnetic field strength as $B = B_0 \exp(\Omega t)$\footnote{The induction equation $\partial B / \partial t = \nabla \times ( {\bf v \times B})$ can be simplified as $\partial B / \partial t = (v B)/L = \Omega B$ where $v$ and $L$ are the typical timescale and typical length scale, and $\Omega$ is the typical angular velocity which should correspond to the angular velocity of the spin or orbital motion of protostars. Assuming $\Omega$ being constant and integrating the simplified equation ($dB/dt=\Omega B$), we have $B = B_0 \exp\,(\Omega t)$, where $B_0$ is an integral constant. This simplification or estimation holds as long as the magnetic feedback can be ignored. The angular velocity $\Omega$ decreases due to the magnetic tension force when the magnetic field (or Lorentz force) is strong. Since we initially assumed very weak fields, exponential growth can occur.} (Figure~\ref{f4}).
During the first $100\,$yr, the magnetic field strength in the region of about $100\,$au around the primary protostar is amplified to $|B|\sim10^{-4}\,$G (Figure~\ref{f5}(a)).
The peak at $R \sim 60\,$au corresponds to the magnetic field amplified by another core.
This amplified magnetic field is strong enough to affect the gas dynamics and decrease the number of dense cores.
Figure~\ref{f4}(b) shows two density ranges where the magnetic field is noticeably amplified: high density region in the circumstellar disk ($n\geq10^{13}\,\cc$) and low density region just above the disk ($n\sim10^{11}\,\cc$).
The rotation motion of the dense cores amplifies the magnetic field in the disk, which in turn amplifies the low-density region above (see also the bottom panels of Figure~\ref{f3}).
The presence of two peaks is transient and becomes less noticeable over time as the magnetic field amplifies throughout the region at the disk's outer edge.

We continue the simulations until $t_{\rm ps} = 500\,$yr.
The amplified ``seed'' magnetic field increases the amplification rate according to the induction equation, $\partial B / \partial t = \nabla \times (\bm{v} \times \bm{B})$, in which the amplified seed field $\bm{B}$ contributes to further increase of the magnetic field.
As a result, the amplified region of the magnetic field strength extends outwards due to the orbital motion of protostars and the rotation of accreting gas (Figures~\ref{f4}(c) and \ref{f5}).

\begin{figure}[t]
\begin{center}
\includegraphics[width=0.74\linewidth]{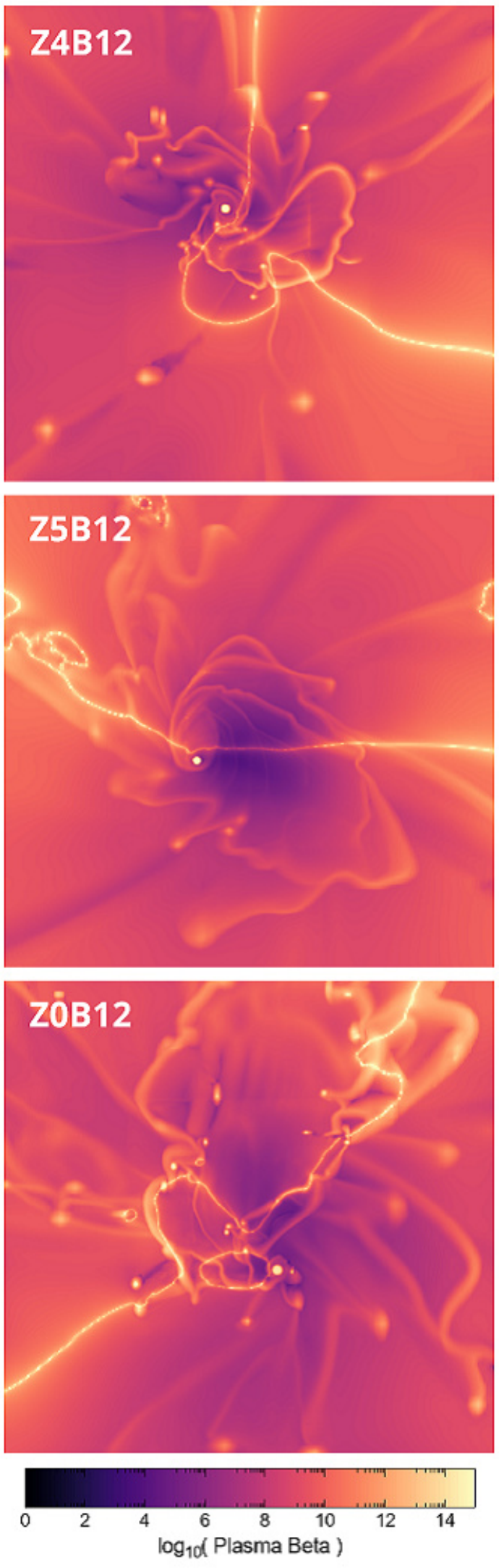}
\end{center}
\caption{
Cross-sectional views of plasma beta on the $z = 0$ planes around the most massive protostar for models Z4B12 (at $t_{\rm ps} = 500$\,yr), Z5B12 (at $t_{\rm ps} = 300$\,yr), and Z0B12 (at $t_{\rm ps} = 500$\,yr). The box sizes of the cross-sectional views are $2000$\,au.
}
\label{f6}
\end{figure}

To clarify the importance of the magnetic field, we plot the plasma beta (ratio of thermal to magnetic pressure) in Figures~\ref{f4}(d)-(f) and \ref{f6}.
Figure~\ref{f6} shows that the plasma beta on the equatorial plane is $\sim\!10^2-10^3$ around the center, indicating that the magnetic pressure is $\sim\!0.1-1\%$ of the thermal pressure (see also Figures~\ref{f4}d-f).
Although the magnetic pressure is lower than the thermal pressure, it can affect the gas dynamics.

The magnetic effects in the magnetized model enhance the ``supercompetitive accretion'' more than in the unmagnetized model.
The angular momentum transfer from the accreting gas increases the mass accretion rate and promotes fragmentation.
Then, the transfer of the orbital angular momentum of the number of protostars facilitates the coalescence of low-mass protostars to the primary one.
The more protostars form, the more the rotational motion amplifies the local magnetic field.
Thus, the magnetic effects positively affect the growth of an SMS in this case.

\begin{figure}[t]
\begin{center}
\includegraphics[width=1.00\linewidth]{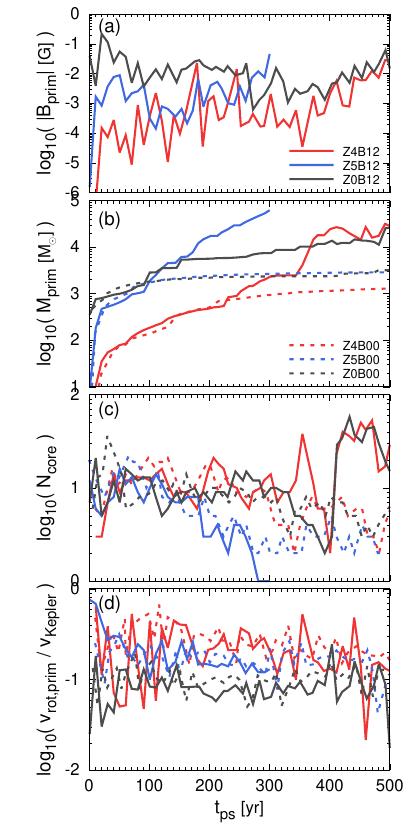}
\end{center}
\caption{
Time evolution of the dense core properties for models Z4B12 (red solid), Z4B00 (red dashed), Z5B12 (blue solid), Z5B00 (blue dashed), Z0B12 (black solid), and Z0B00 (black dashed), respectively.
The solid lines show the magnetized models, whereas the dashed lines show the unmagnetized models.
Panels: (a) absolute magnetic field strength, (b) mass of the primary core, (c) number of living dense cores (protostars), and (d) ratio of rotational velocity to the Keplerian velocity of the primary core.
}
\label{f7}
\end{figure}

\begin{figure*}[t]
\begin{center}
\includegraphics[width=1.0\linewidth]{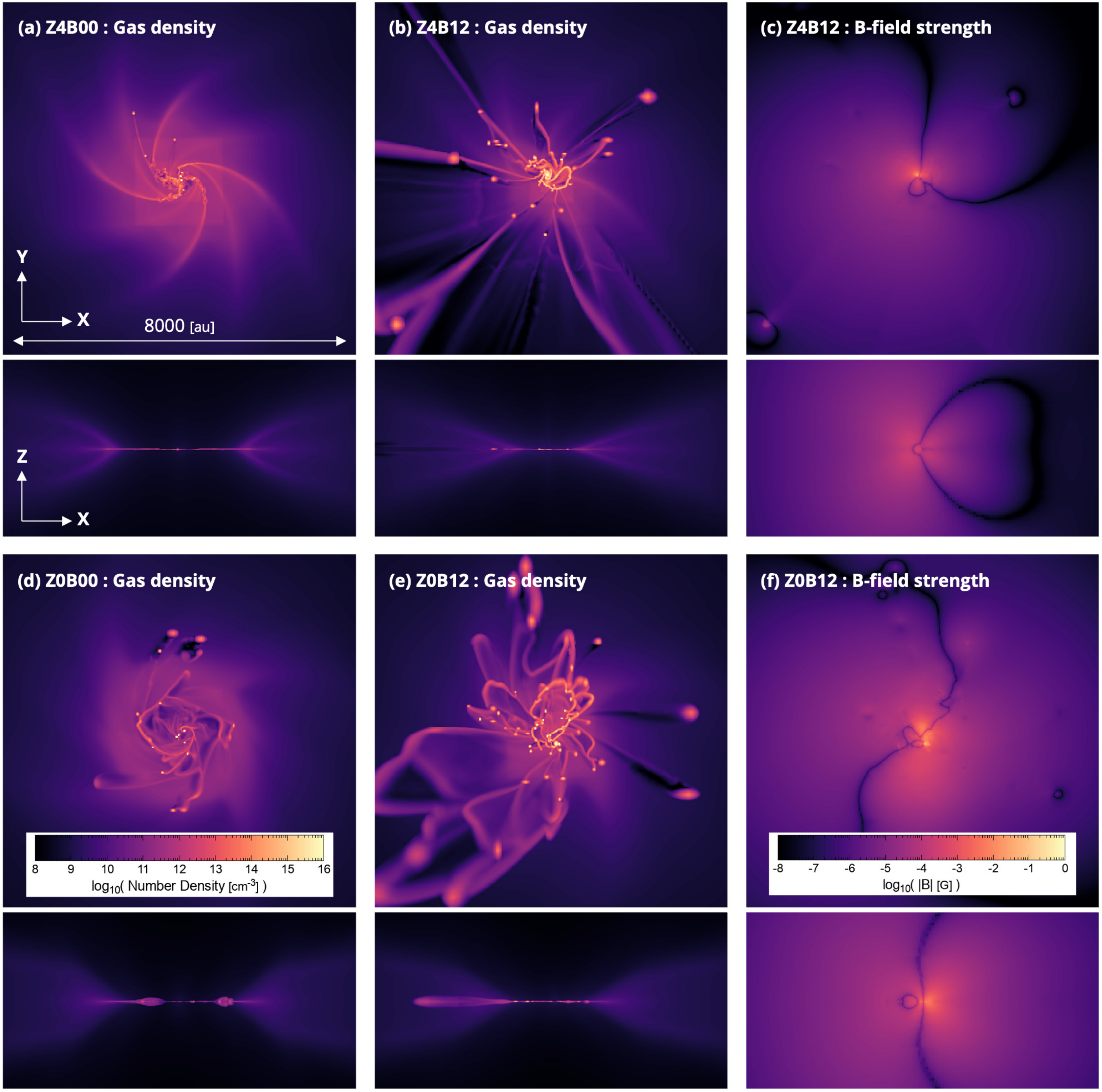}
\end{center}
\caption{
Cross-sectional views on the $z = 0$ and $y = 0$ planes around the most massive protostar at $t_{\rm ps}=500\,$yr after the first protostar formation.
Panels: (a, b, d, and e) gas number density for models Z4B00, Z4B12, Z0B00, and Z0B12, (c and f) absolute magnetic field strength for models Z4B12 and Z0B12.
}
\label{f8}
\end{figure*}

\begin{figure*}[t]
\begin{center}
\includegraphics[width=0.95\linewidth]{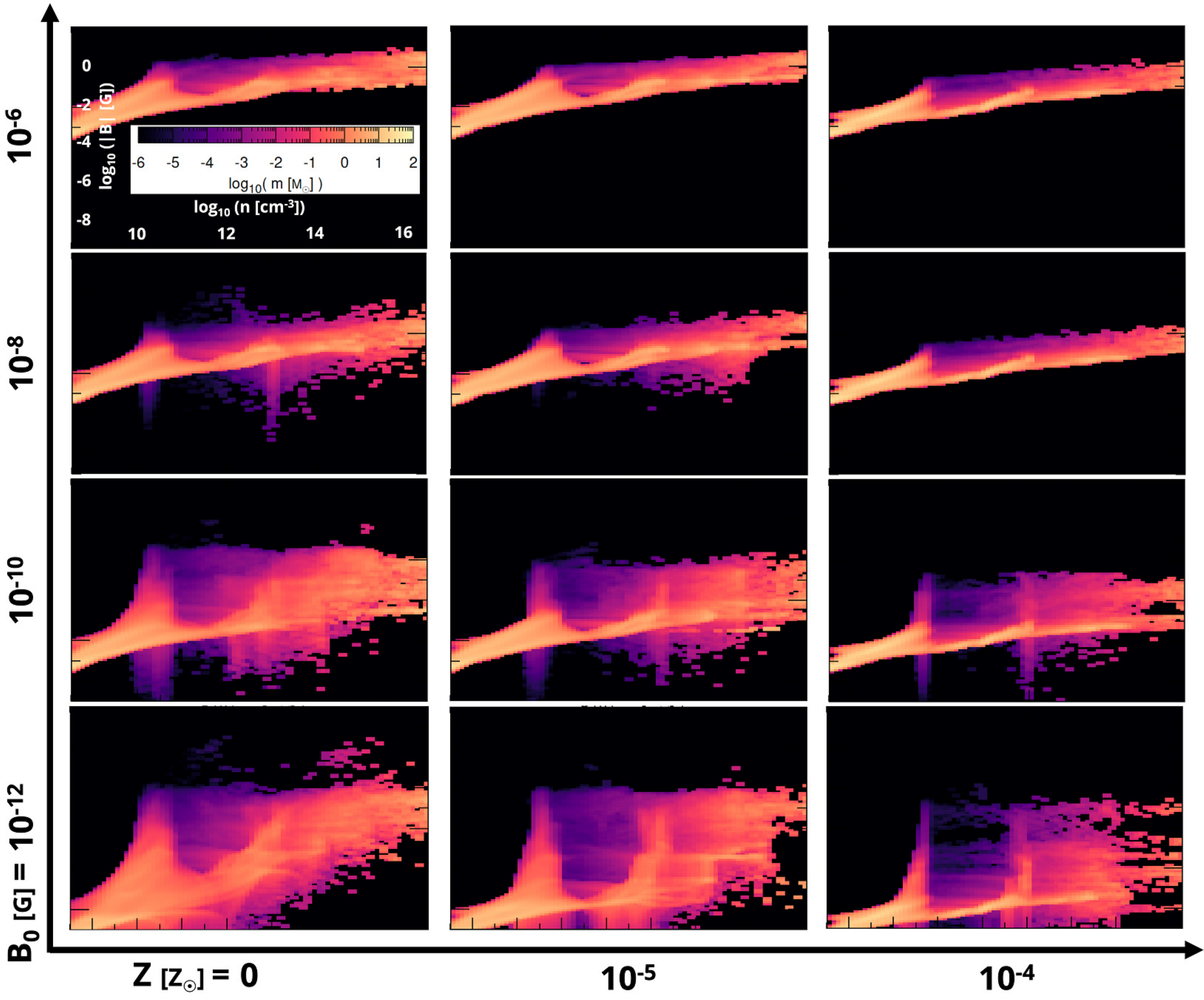}
\end{center}
\caption{
Phase diagrams of the absolute magnetic field strength for all models at $t_{\rm ps} = 100\,$yr after the first protostar formation.
}
\label{f9}
\end{figure*}

\begin{figure*}[t]
\begin{center}
\includegraphics[width=0.95\linewidth]{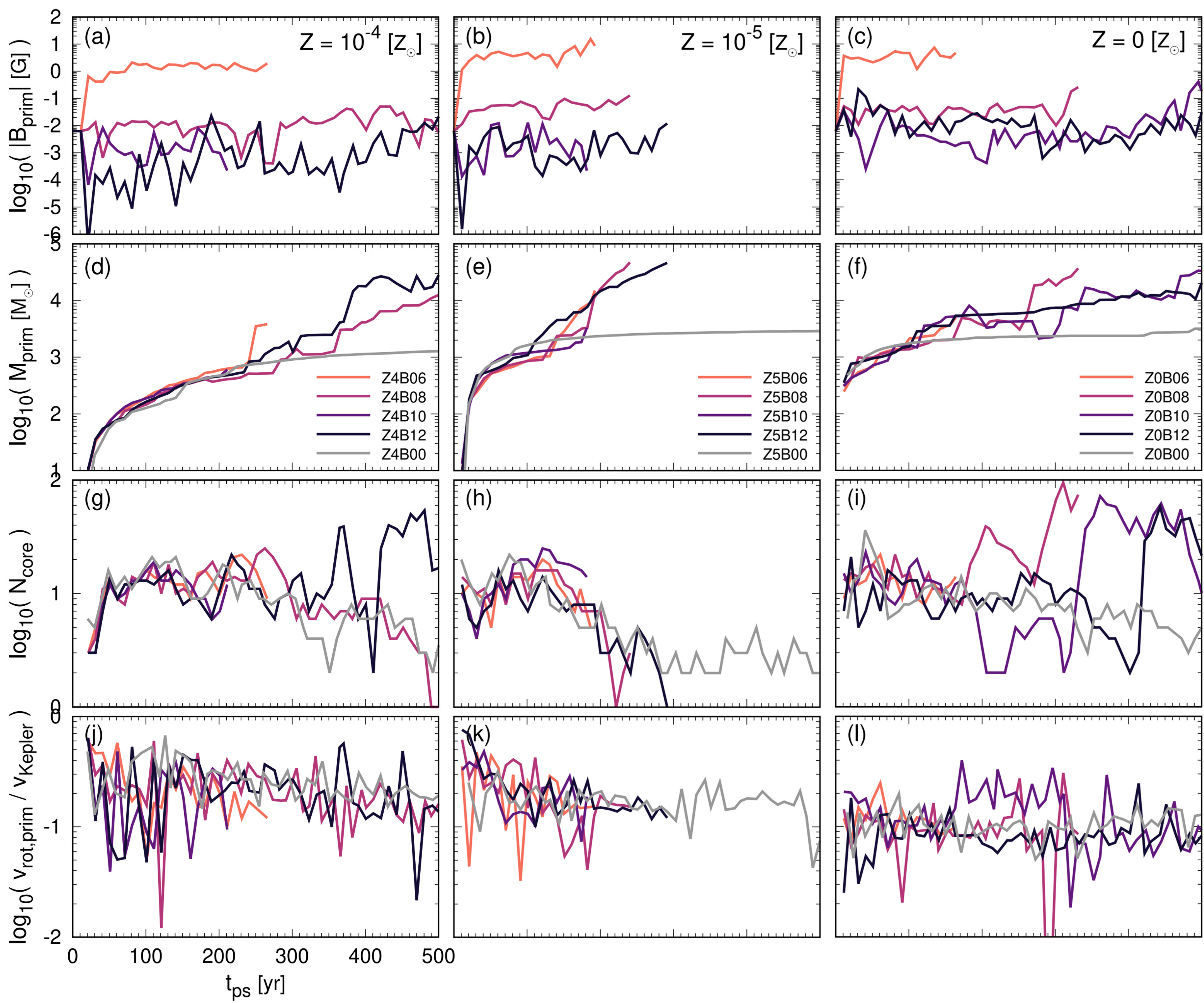}
\end{center}
\caption{
Same as Figure~\ref{f7} but for all models.
}
\label{f10}
\end{figure*}

Figure~\ref{f7}(b) compares the time evolution of the primary protostellar mass between the magnetized (Z4B12; red solid line) and unmagnetized (Z4B00; red dashed line) models.
The early mass growth histories are almost identical until $t_{\rm ps}\sim240\,$yr.
After the epoch, the mass growth accelerates for the magnetized model compared with the unmagnetized one.
In the magnetized model, the mass of the primary star suddenly rises due to the runaway coalescence of almost all low-mass protostars (Figure~\ref{f7}(c)) at $t_{\rm ps}\sim400\,$yr.
After that, the number of protostars increases to $N_{\rm core}\sim50$ due to the high mass accretion rate.
The intermittent fragmentation and coalescence allow the primary star to gain mass more efficiently.

\subsection{Dependence on the metallicity} \label{sec:res-Z}

Next, we examine the dependence of the magnetic effects on metallicity ($Z$).
The gas temperature decreases as the metallicity increases.
Thus the mass accretion rate should be lower in the cloud with higher metallicity \citep[e.g.,][]{hosokawa09}.
The unmagnetized models (dashed lines in Figure~\ref{f7}(b)) represent this trend as the mass growth rate of the primary protostar decreases with metallicity.
At the end of the simulations $t_{\rm ps} = 500\,$yr, the masses of primary protostars reach about a few $10^3\,\msun$, which are consistent with the previous unmagnetized simulation \cite[see Figure~4(a) of][]{chon20}.

By considering the magnetic effects (solid lines of Figure~\ref{f7}(b)), the masses of the primary protostars at $t_{\rm ps} = 500\,$yr become about $10$ times greater than those in the unmagnetized models for all metallicity models examined.
Note that the mass infall rate far from the central region is almost the same between the magnetized and unmagnetized models. On the other hand, the mass infall rate near the center is higher in the magnetized models than in the unmagnetized models. Thus, the magnetic field affects the mass accretion rate near the center.
\footnote{
It should be noted that we confirmed that some models show a high mass accretion rate $\sim\!100\,\msunyr$. However, we calculated the cloud evolution only for $\lesssim\! 500$\,yr after the first protostar formation. We need further calculations to evaluate the long-term mass accretion rate correctly.
}
At the early accretion phase ($t_{\rm ps} < 20\,$yr), the amplification rate of the magnetic field strength is higher with decreasing metallicity (Figure~\ref{f7}(a)).
The origin of this early difference more clearly appears in the ``seed'' magnetic field amplified by the spin of protostars (Figure~\ref{f7}(a)), as the lower metallicity causes a higher accretion rate and the central high-density gas region fragments into many protostars (Figure~\ref{f7}(c)).
After that, all magnetized models will eventually undergo significant magnetic field amplification.
The angular momentum transport by the amplified magnetic field promotes gas accretion and protostellar mergers.
After only $500\,$yr, the primary protostars obtain a mass $\sim\!10^4\,\msun$.

The mass growth rate is highest in the extremely low metallicity model (Z5B12) rather than the metal-free model (Z0B12), as shown in Figure~\ref{f7}(b).
The increase in metallicity has both positive and negative effects on mass growth: increasing the number of protostars which contributes to exponential magnetic field amplification, and decreasing the mass accretion rate due to the lower temperature of the gas cloud.
The mass growth history of the primary protostar is almost similar between the unmagnetized models Z5B00 and Z0B00 (Figure~\ref{f7}(b)).
Therefore, we find that the positive feedback effect wins out at the very low metallicity value $Z/\zsun = 10^{-5}$, but as the metallicity increases, the negative feedback becomes as important.
We see no significant further net feedback effect.

Figure~\ref{f8} shows 2D maps of density and magnetic field strength around the primary protostar at the end of the simulation ($t_{\rm ps} = 500\,$yr).
In the magnetized models, there are many protostars formed by the fragmentation burst at $t_{\rm ps} = 400\,$yr (Figure~\ref{f7}(c)).
Some protostars are escaped farther away from the cloud center (the primary protostar) in the magnetized models (Figures~\ref{f8}(b) and (e)) than in the unmagnetized models (Figures~\ref{f8}(a) and (d)).
This is due to the increased efficiency of $N$-body interactions since the angular momentum transport due to magnetic effects causes more protostars to fall into the cloud's central region.
We conclude that the rapid accreting gas flow moves protostars away from the cloud's center in the magnetized models rather than in the unmagnetized models.

The structures of the infalling-accreting envelopes in edge-on views ($y = 0$ planes) depend on two parameters.
The envelope height decreases with metallicity by decreasing temperature or thermal pressure.
The inner diameter of the envelope (or the diameter of the thin disk) decreases with increasing magnetic field strength due to the angular momentum transfer.

\subsection{Dependence on the initial magnetic field strength} \label{sec:res-B0}

Finally, we examine the dependence on the initial magnetic field strength ($B_0$).
Figure~\ref{f9} summarizes $B$-$n$ diagrams at the same epoch ($t_{\rm ps} = 100\,$yr) for all models.
We confirm the amplification of the magnetic field in all models.
There are two distinct density ranges of the amplification: high density region on the circumstellar disk ($n\geq10^{13}\,\cc$) and low density region above the disk ($n\sim10^{11}\,\cc$) (as explained in Section~\ref{sec:res-fiducial}).

Figure~\ref{f10} summarizes the time evolution of the protostar properties.
The mass growth is faster in the magnetized models than in the unmagnetized models.
However, the mass growth rate does not always increase with $B_0$.
The gas infall (or negative radial) velocity to the center is large in the amplified region due to the efficient removal of the angular momentum by the magnetic field. 
The infall velocity, and hence the mass growth rate of the protostar, is naively considered to increase with the initial magnetic field strength. However, the opposite results are often observed, in which the stronger initial magnetic field slows protostellar mass growth (Figures~\ref{f10}\,d-f). 
This is because the rotation-driven amplification of the magnetic field  (Figure~\ref{f1}) does not proceed uniformly throughout the cloud, but proceeds first around the center of the gas cloud where the rotation speed is fastest.

The strong magnetic field region spreads to the outer area where the rotation speed is slower (Figure~\ref{f5}a).
Efficient gas accretion occurs when the gas and dense cores are sufficiently present in the amplification region. 
If most of the gas (and other cores) in the amplified region accretes onto the primary core, the rotation-driven amplification of the magnetic field would not work well because of the deficit of the accreting gas. 
In this case, the magnetic field amplification stagnates until the gas is supplied from the area outside the amplification region. 
Figures~10(g)-(i) show strong oscillations in the number of cores, especially in magnetized models, indicating violent fragmentation following the rapid coalescence of cores. 
Figures~10(f) and (i) show the synchronous change in mass growth rate and the number of cores.

Figures~\ref{f10}(d)-(f) confirm that the mass growth is maximized with $Z/\zsun = 10^{-5}$ for the different values of $B_0$.

Figures~\ref{f10}(j)-(l) also present that the rotational velocities stay low enough that the primary star attains a rotational velocity that is about $0.1$ times the Keplerian velocity, regardless of the model parameters.
This is consistent with the requirement for SMS to have low surface velocities of only $10$--$20$\% of the Keplerian value to continue accreting and reach their final masses \citep{haemmerle18}.
In addition, the figures indicate that the magnetic field has a subdominant role in determining the rotation of the protostar.

\section{Discussion} \label{sec:dis}

\subsection{Evaluation of the final stellar mass}

We represent that the magnetic effects enhance the mass growth rate in the metal-enriched ACHs.
One of the limitations of this study is that the mass of a dense core with $n \geq n_{\rm th}$ is the upper limit for the protostellar mass formed inside the dense core.
We have to confirm whether a large amount of gas can accrete efficiently to the protostar in future work.
We note that the rotation-driven amplification of the magnetic field is more efficient inside the unresolved region of this study, $n > n_{\rm th} = 10^{16}\,\cc$ because the angular velocity ($\Omega = v/r$) increases with decreasing the distance from the central protostar.
Then the magnetic effects could also support the efficient mass accretion onto the protostar in future work.

In addition, the bursts of fragmentation and merger of protostars also promote massive star formation by preventing the protostellar radiative feedback.
Figure~4(d) of \cite{chon20} showed that about $3000$ stars formed during $10^4\,$yr after the first protostar.
The magnetic effects enhance the ``supercompetitive accretion'' and causes some fraction (or all) of stars to fall on the primary star.
Such a highly variable mass accretion rate is advantageous for the DCBH formation because the frequent periods of high accretion rate can inflate the protostar and keep the effective temperature low enough to suppress the radiation feedback that can cease the mass accretion onto the protostar \citep[e.g.,][]{sakurai16}.

\subsection{Mass growth from IMBH to SMBH}

After forming the IMBHs in the metal-enriched ACHs, the issue is whether a high accretion rate can be realized to grow from IMBHs to SMBHs.
\cite{yajima17} and \cite{park22arxiv} investigated the super-Eddington accretion onto IMBHs in low-metallicity environments.
\cite{regan20} showed that $2/3$ of the low-metallicity ACHs ($Z/\zsun \geq 10^{-3}$) have a high accretion rate ($\geq\!0.1\,\msunyr$) by analyzing the Renaissance simulation data.
Recently, \cite{chon21} reported the metallicity criterion for DCBH-SMBH growth as $Z/\zsun \sim 10^{-5}$--$10^{-3}$.
In this way, there is a growing chance for the DCBHs to become SMBHs in low-metallicity environments.

\subsection{Non-ideal MHD effects}\label{sec:dis-caution}

We calculate the ideal MHD equations in this study.
If the non-ideal MHD effects become efficient and significantly dissipate the magnetic field within the disk, the rotation-driven amplification would not be efficient.
\cite{higuchi18} presented that the non-ideal MHD effects are ineffective in the metal-free star-forming clouds.
We can assume that the non-ideal MHD effects are also ineffective in the metal-free ACHs because the gas temperature is always higher in the metal-free ACHs than in the typical metal-free clouds (with $\sim\!10^3\,\msun$) at any stage of star formation with different gas densities.
On the other hand, no one has yet examined the non-ideal MHD effects on metal-enriched ACHs.
This is one of the future works.

In the metal-enriched ACHs with $Z/\zsun = 10^{-5}$ and $10^{-4}$, we assume that the non-ideal MHD effects are negligible because the gas temperature, which minimizes at about $200$ and $100\,$K (Figure~\ref{f2}), respectively, is hot enough.
The more metal-enriched ACHs with $Z/\zsun = 10^{-3}$ cool below to $100\,$K \citep[Figure~1 of][]{chon20}.
In such a low-temperature cloud, the non-ideal MHD effects could work.
In addition, the mass growth rate in the unmagnetized ACH with $Z/\zsun = 10^{-3}$ is about an order of magnitude smaller than that in one with $Z/\zsun \leq 10^{-4}$ \citep{chon20}.
This parameter ($Z/\zsun = 10^{-3}$) does not seem to lead to the IMBH formation.

\subsection{Numerical Resolution}\label{sec:dis-resolution}

As described in Section~\ref{sec:methodology}, we resolve the Jeans wavelength with 32 cells to generate a new finer grid. 
However, we relaxed the condition to resolve the Jeans wavelength after generating the highest level grid. 
In the finest grid, the Jeans length was not sufficiently resolved in a high-density region. 
We need about ten times higher spatial resolution to resolve the Jeans length.
Thus, we may overestimate/underestimate the number of fragments. 
Note that when the massive core or protostar exits, we may need to other criteria to resolve fragmentation \citep[see Equations. (13)--(18) of][]{matsumoto03}.

Most of the large or massive fragments in the simulation are reasonably resolved, with about $10$ cells inside the large fragments. 
However, smaller fragments are not well resolved and may be transient in the simulation.  
The high-resolution simulations resolving small fragments or clumps are extremely demanding computationally.

We expect that the number of protostars and amplification timescale should be changed with high-resolution simulations since small fragments may be resolved and therefore survive in the simulation. 
However, we also expect the amplification of the magnetic field shown in this study is not qualitatively changed as long as the orbital and spin motion of protostars are resolved. 
The spatial resolution should be improved and we can address this fully in future studies.

\section{Conclusion} \label{sec:con}

We confirm the rotation-driven amplification mechanism of the magnetic field during the accretion phase of the DCBH formation in metal-enriched ACHs.
Although increasing the metallicity reduces the mass accretion, the collapsing central region is gravitationally unstable and fragments to form many protostars.
The protostars drive magnetic field amplification to promote subsequent gas accretion and star coalescence.
The magnetic amplification effect is efficient independent of the initial magnetic field strength in the dynamically weak initial magnetic field regime we study.
The amplification mechanism of the magnetic field works even in the metal-enriched ACHs with the primordial magnetic field strength $B_0/\,{\rm G} = 10^{-12}$ at $\nh = 10^4\,\cc$.
The mass growth rate is maximal for the extremely metal-poor ACHs with $Z/\zsun = 10^{-5}$, which is a qualitatively different trend from the unmagnetized simulations.
We conclude that the exponential magnetic field amplification is realized in the metal-enriched ACHs and relaxes the DCBH formation criterion.

\begin{acknowledgments}
This work used the computational resources of the HPCI system provided by the supercomputer system SX-Aurora TSUBASA at Tohoku University Cyber Sciencecenter and Osaka University Cybermedia Center through the HPCI System Research Project (Project ID: hp210004 and hp220003) and Earth Simulator at JAMSTEC provided by 2021 and 2022 Koubo Kadai.
Numerical analyses were carried out on analysis servers at Center for Computational Astrophysics, National Astronomical Observatory of Japan.
S.H. was supported by JSPS KAKENHI Grant Numbers JP18H05222, JP21H01123, and JP21K13960 and Qdai-jump Research Program 02217.
M.N.M. was supported by JSPS KAKENHI Grant Numbers JP17H06360, JP17K05387, JP17KK0096, JP21K03617, and JP21H00046 and University Research Support Grant 2019 from the National Astronomical Observatory of Japan (NAOJ).
S.B. was supported by a Discovery Grant from NSERC.
\end{acknowledgments}




\bibliography{ms}{}

\begin{thebibliography}{}
\expandafter\ifx\csname natexlab\endcsname\relax\def\natexlab#1{#1}\fi
\providecommand{\url}[1]{\href{#1}{#1}}
\providecommand{\dodoi}[1]{doi:~\href{http://doi.org/#1}{\nolinkurl{#1}}}
\providecommand{\doeprint}[1]{\href{http://ascl.net/#1}{\nolinkurl{http://ascl.net/#1}}}
\providecommand{\doarXiv}[1]{\href{https://arxiv.org/abs/#1}{\nolinkurl{https://arxiv.org/abs/#1}}}

\bibitem[{{Agarwal} {et~al.}(2012){Agarwal}, {Khochfar}, {Johnson}, {Neistein},
  {Dalla Vecchia}, \& {Livio}}]{agarwal12}
{Agarwal}, B., {Khochfar}, S., {Johnson}, J.~L., {et~al.} 2012, \mnras, 425,
  2854, \dodoi{10.1111/j.1365-2966.2012.21651.x}

\bibitem[{{Ba{\~n}ados} {et~al.}(2016){Ba{\~n}ados}, {Venemans}, {Decarli},
  {Farina}, {Mazzucchelli}, {Walter}, {Fan}, {Stern}, {Schlafly}, {Chambers},
  {Rix}, {Jiang}, {McGreer}, {Simcoe}, {Wang}, {Yang}, {Morganson}, {De Rosa},
  {Greiner}, {Balokovi{\'c}}, {Burgett}, {Cooper}, {Draper}, {Flewelling},
  {Hodapp}, {Jun}, {Kaiser}, {Kudritzki}, {Magnier}, {Metcalfe}, {Miller},
  {Schindler}, {Tonry}, {Wainscoat}, {Waters}, \& {Yang}}]{banados16}
{Ba{\~n}ados}, E., {Venemans}, B.~P., {Decarli}, R., {et~al.} 2016, \apjs, 227,
  11, \dodoi{10.3847/0067-0049/227/1/11}

\bibitem[{{Chon} {et~al.}(2021){Chon}, {Hosokawa}, \& {Omukai}}]{chon21}
{Chon}, S., {Hosokawa}, T., \& {Omukai}, K. 2021, \mnras, 502, 700,
  \dodoi{10.1093/mnras/stab061}

\bibitem[{{Chon} \& {Omukai}(2020)}]{chon20}
{Chon}, S., \& {Omukai}, K. 2020, \mnras, 494, 2851,
  \dodoi{10.1093/mnras/staa863}

\bibitem[{{Haemmerl{\'e}} {et~al.}(2018){Haemmerl{\'e}}, {Woods}, {Klessen},
  {Heger}, \& {Whalen}}]{haemmerle18}
{Haemmerl{\'e}}, L., {Woods}, T.~E., {Klessen}, R.~S., {Heger}, A., \&
  {Whalen}, D.~J. 2018, \apjl, 853, L3, \dodoi{10.3847/2041-8213/aaa462}

\bibitem[{{Higuchi} {et~al.}(2018){Higuchi}, {Machida}, \& {Susa}}]{higuchi18}
{Higuchi}, K., {Machida}, M.~N., \& {Susa}, H. 2018, \mnras, 475, 3331,
  \dodoi{10.1093/mnras/sty046}

\bibitem[{{Hirano} \& {Bromm}(2017)}]{hirano17}
{Hirano}, S., \& {Bromm}, V. 2017, \mnras, 470, 898,
  \dodoi{10.1093/mnras/stx1220}

\bibitem[{{Hirano} {et~al.}(2017){Hirano}, {Hosokawa}, {Yoshida}, \&
  {Kuiper}}]{hirano17sv}
{Hirano}, S., {Hosokawa}, T., {Yoshida}, N., \& {Kuiper}, R. 2017, Science,
  357, 1375, \dodoi{10.1126/science.aai9119}

\bibitem[{{Hirano} \& {Machida}(2022)}]{hirano22}
{Hirano}, S., \& {Machida}, M.~N. 2022, \apjl, 935, L16,
  \dodoi{10.3847/2041-8213/ac85e0}

\bibitem[{{Hirano} {et~al.}(2021){Hirano}, {Machida}, \& {Basu}}]{hirano21}
{Hirano}, S., {Machida}, M.~N., \& {Basu}, S. 2021, \apj, 917, 34,
  \dodoi{10.3847/1538-4357/ac0913}

\bibitem[{{Hosokawa} \& {Omukai}(2009)}]{hosokawa09}
{Hosokawa}, T., \& {Omukai}, K. 2009, \apj, 703, 1810,
  \dodoi{10.1088/0004-637X/703/2/1810}

\bibitem[{{Hosokawa} {et~al.}(2012){Hosokawa}, {Omukai}, \&
  {Yorke}}]{hosokawa12}
{Hosokawa}, T., {Omukai}, K., \& {Yorke}, H.~W. 2012, \apj, 756, 93,
  \dodoi{10.1088/0004-637X/756/1/93}

\bibitem[{{Inayoshi} {et~al.}(2020){Inayoshi}, {Visbal}, \&
  {Haiman}}]{inayoshi20}
{Inayoshi}, K., {Visbal}, E., \& {Haiman}, Z. 2020, \araa, 58, 27,
  \dodoi{10.1146/annurev-astro-120419-014455}

\bibitem[{{Inayoshi} {et~al.}(2015){Inayoshi}, {Visbal}, \&
  {Kashiyama}}]{inayoshi15}
{Inayoshi}, K., {Visbal}, E., \& {Kashiyama}, K. 2015, \mnras, 453, 1692,
  \dodoi{10.1093/mnras/stv1654}

\bibitem[{{Latif} {et~al.}(2013){Latif}, {Schleicher}, {Schmidt}, \&
  {Niemeyer}}]{latif13}
{Latif}, M.~A., {Schleicher}, D.~R.~G., {Schmidt}, W., \& {Niemeyer}, J. 2013,
  \mnras, 433, 1607, \dodoi{10.1093/mnras/stt834}

\bibitem[{{Latif} {et~al.}(2022){Latif}, {Whalen}, {Khochfar}, {Herrington}, \&
  {Woods}}]{latif22}
{Latif}, M.~A., {Whalen}, D.~J., {Khochfar}, S., {Herrington}, N.~P., \&
  {Woods}, T.~E. 2022, \nat, 607, 48, \dodoi{10.1038/s41586-022-04813-y}

\bibitem[{{Machida} \& {Doi}(2013)}]{machida13}
{Machida}, M.~N., \& {Doi}, K. 2013, \mnras, 435, 3283,
  \dodoi{10.1093/mnras/stt1524}

\bibitem[{{Machida} \& {Nakamura}(2015)}]{machida15}
{Machida}, M.~N., \& {Nakamura}, T. 2015, \mnras, 448, 1405,
  \dodoi{10.1093/mnras/stu2633}

\bibitem[{{Matsumoto} \& {Hanawa}(2003)}]{matsumoto03}
{Matsumoto}, T., \& {Hanawa}, T. 2003, \apj, 595, 913, \dodoi{10.1086/377367}

\bibitem[{{McKee} {et~al.}(2020){McKee}, {Stacy}, \& {Li}}]{mckee20}
{McKee}, C.~F., {Stacy}, A., \& {Li}, P.~S. 2020, \mnras, 496, 5528,
  \dodoi{10.1093/mnras/staa1903}

\bibitem[{{Nagele} {et~al.}(2022){Nagele}, {Umeda}, {Takahashi}, {Yoshida}, \&
  {Sumiyoshi}}]{nagele22arXiv}
{Nagele}, C., {Umeda}, H., {Takahashi}, K., {Yoshida}, T., \& {Sumiyoshi}, K.
  2022, \mnras, 517, 1584, \dodoi{10.1093/mnras/stac2495}

\bibitem[{{Omukai}(2001)}]{omukai01}
{Omukai}, K. 2001, \apj, 546, 635, \dodoi{10.1086/318296}

\bibitem[{{Omukai} {et~al.}(2008){Omukai}, {Schneider}, \& {Haiman}}]{omukai08}
{Omukai}, K., {Schneider}, R., \& {Haiman}, Z. 2008, \apj, 686, 801,
  \dodoi{10.1086/591636}

\bibitem[{{Park} {et~al.}(2022){Park}, {Chiaki}, \& {Wise}}]{park22arxiv}
{Park}, K., {Chiaki}, G., \& {Wise}, J.~H. 2022, \apj, 936, 116,
  \dodoi{10.3847/1538-4357/ac886c}

\bibitem[{{Regan} {et~al.}(2020){Regan}, {Haiman}, {Wise}, {O'Shea}, \&
  {Norman}}]{regan20}
{Regan}, J.~A., {Haiman}, Z., {Wise}, J.~H., {O'Shea}, B.~W., \& {Norman},
  M.~L. 2020, The Open Journal of Astrophysics, 3, E9,
  \dodoi{10.21105/astro.2006.14625}

\bibitem[{{Sakurai} {et~al.}(2016){Sakurai}, {Vorobyov}, {Hosokawa}, {Yoshida},
  {Omukai}, \& {Yorke}}]{sakurai16}
{Sakurai}, Y., {Vorobyov}, E.~I., {Hosokawa}, T., {et~al.} 2016, \mnras, 459,
  1137, \dodoi{10.1093/mnras/stw637}

\bibitem[{{Sur} {et~al.}(2010){Sur}, {Schleicher}, {Banerjee}, {Federrath}, \&
  {Klessen}}]{sur10}
{Sur}, S., {Schleicher}, D.~R.~G., {Banerjee}, R., {Federrath}, C., \&
  {Klessen}, R.~S. 2010, \apjl, 721, L134, \dodoi{10.1088/2041-8205/721/2/L134}

\bibitem[{{Tanaka} \& {Li}(2014)}]{tanaka14}
{Tanaka}, T.~L., \& {Li}, M. 2014, \mnras, 439, 1092,
  \dodoi{10.1093/mnras/stu042}

\bibitem[{{Umeda} {et~al.}(2016){Umeda}, {Hosokawa}, {Omukai}, \&
  {Yoshida}}]{umeda16}
{Umeda}, H., {Hosokawa}, T., {Omukai}, K., \& {Yoshida}, N. 2016, \apjl, 830,
  L34, \dodoi{10.3847/2041-8205/830/2/L34}

\bibitem[{{Wise} {et~al.}(2019){Wise}, {Regan}, {O'Shea}, {Norman}, {Downes},
  \& {Xu}}]{wise19}
{Wise}, J.~H., {Regan}, J.~A., {O'Shea}, B.~W., {et~al.} 2019, \nat, 566, 85,
  \dodoi{10.1038/s41586-019-0873-4}

\bibitem[{{Woods} {et~al.}(2019){Woods}, {Agarwal}, {Bromm}, {Bunker}, {Chen},
  {Chon}, {Ferrara}, {Glover}, {Haemmerl{\'e}}, {Haiman}, {Hartwig}, {Heger},
  {Hirano}, {Hosokawa}, {Inayoshi}, {Klessen}, {Kobayashi}, {Koliopanos},
  {Latif}, {Li}, {Mayer}, {Mezcua}, {Natarajan}, {Pacucci}, {Rees}, {Regan},
  {Sakurai}, {Salvadori}, {Schneider}, {Surace}, {Tanaka}, {Whalen}, \&
  {Yoshida}}]{woods19}
{Woods}, T.~E., {Agarwal}, B., {Bromm}, V., {et~al.} 2019, \pasa, 36, e027,
  \dodoi{10.1017/pasa.2019.14}

\bibitem[{{Xu} {et~al.}(2008){Xu}, {O'Shea}, {Collins}, {Norman}, {Li}, \&
  {Li}}]{xu08}
{Xu}, H., {O'Shea}, B.~W., {Collins}, D.~C., {et~al.} 2008, \apjl, 688, L57,
  \dodoi{10.1086/595617}

\bibitem[{{Yajima} {et~al.}(2017){Yajima}, {Ricotti}, {Park}, \&
  {Sugimura}}]{yajima17}
{Yajima}, H., {Ricotti}, M., {Park}, K., \& {Sugimura}, K. 2017, \apj, 846, 3,
  \dodoi{10.3847/1538-4357/aa8269}

\end{thebibliography}
\bibliographystyle{aasjournal}

\end{document}